# Multi-Stage JavaScript

Anthony Savidis[1,2], Yannis Apostolidis[1], Yannis Lilis[1]
[1]Institute of Computer Science, FORTH
[2]Department of Computer Science, University of Crete
{as, japostol, lilis}@ics.forth.gr

## Abstract

Multi-stage languages support generative metaprogramming via macros evaluated in a process preceding the actual interpretation or compilation of the program in which they are used. Macros update the source of their hosting program by emitting code that takes their place in the file, while their code may also be produced, fully or partially, by nested macros. All macros at the same nesting belong to the same stage, with the outer stage collecting the macros affecting only the main program.

We extended JavaScript with staging annotations and implemented them in Spider Monkey, emitting pure JavaScript code as the final outcome of stage computation. We discuss how the original Spider Monkey system is minimally affected with extensions in the syntax, parser and internal AST structures, and the addition of an unparser, a staging loop, some library functions and a debugger backend component for AST inspection. Since stages have a generative metaprogramming role we do not foresee any interplay with the browser DOM, and thus there is no reason to repeat their evaluation on every page load.

Hence, such JavaScript extensions are meant only for development-time, emitting pure JavaScript code that can be run in any browser. Finally, to enable debugging stages in any browser we implemented a pure JavaScript client, communicating with the extended Spider Monkey, and offering the necessary AST display and unparsing that a browser debugger does not provide.

## 1. Introduction

Multi-stage languages [11] support embedded meta-programs that are evaluated in a process preceding the actual interpretation or compilation of the program in which they are used. Usually, the execution of a meta-program evaluates to an Abstract Syntax Tree (ASTs) that takes its place in the original program. Since this procedure is a characteristic of macros systems, the term macro is also used to denote a generative meta-program. The general staging process is depicted under Figure 1 and repeats until no stages exist. Staged code can appear statically, either directly inside the main program or within other stages, or dynamically, in the output of a stage program.

Staging or macro-systems are available for a variety of languages, some through extensions and some as a built-in feature, with typical examples including Common Lisp [7], MetaML [8], MetaOccaml [1], Template Haskell [9], and Converge [12]. When it comes to broadly used languages, a few implementations exist for Java, like Mint [15] and Backstage Java [6], and a couple for JavaScript like [2] and [14]. From the various challenges involved in making a multi-stage language we focused on ease-of-use for the language user and ease-of-implementation for the language author.

To accommodate these needs [5]: (i) the multi-stage language should be a minimal superset of the host language; and (ii) its implementation should not repeat anything already implemented for the host language. Reflecting these goals, we implemented a multi-stage extension of JavaScript on top of the Spider Monkey engine, with a set of small-scale extensions outlined under Figure 2. This effort has resulted in a full-scale multi-stage implementation of JavaScript, while we also added the extra features of the *integrated* model for stage meta-programs introduced in [5]. The latter is as expressive as macros, and includes an extra staging tag to support larger-scale meta-programs.

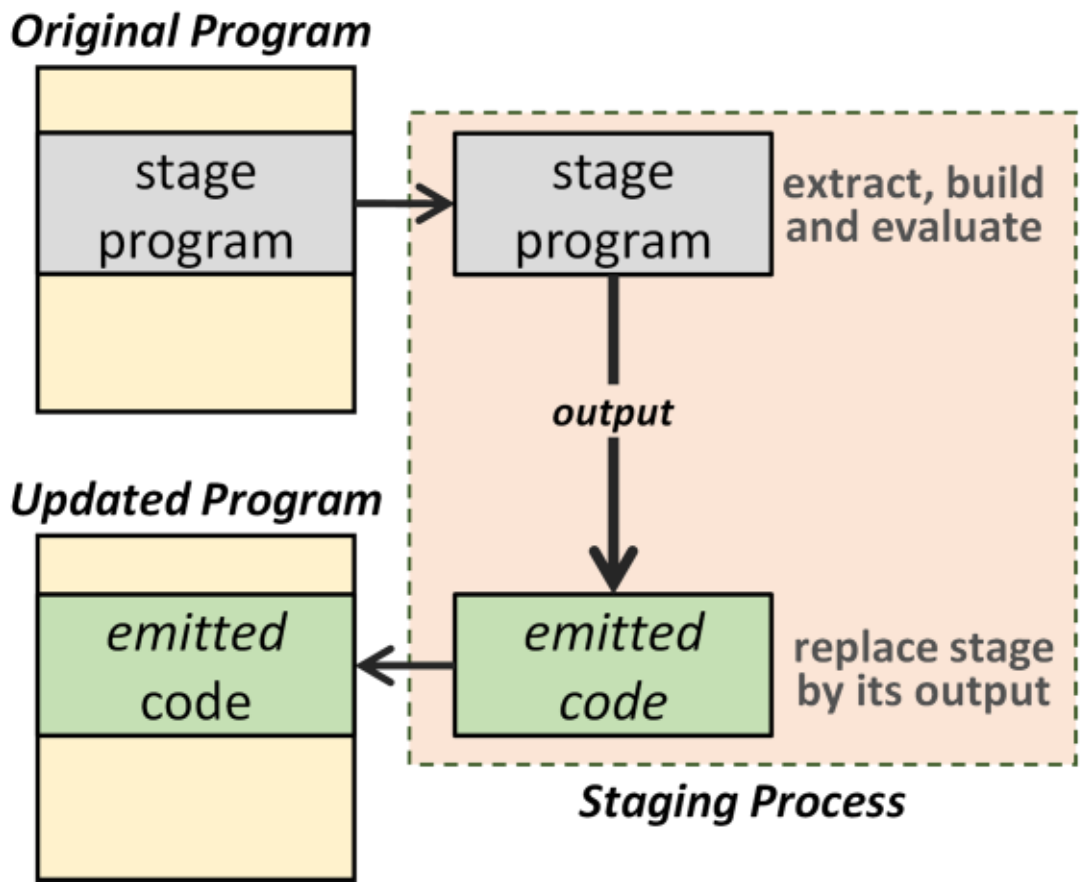

Figure 1 Stage-evaluation process in multi-stage languages.

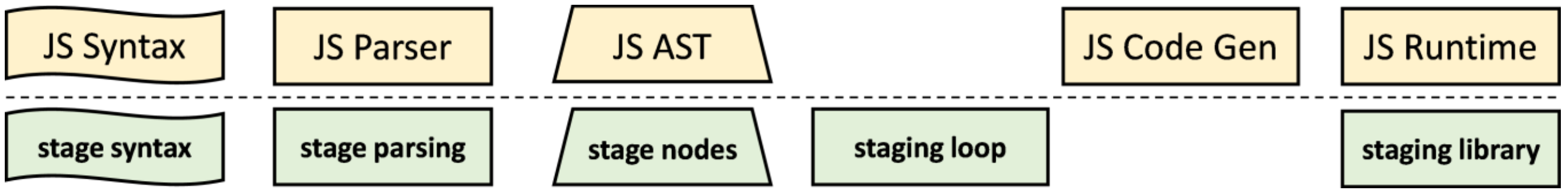

Figure 2 Outline of the original Java Script components in the Spider Monkey implementation (top layer) and the respective extensions we have introduced to support staging (bottom layer).

During testing, we have built comprehensive stage programs, and experimented with generative approaches for web user-interfaces with final JavaScript code entirely produced by stages. Although we could try non-web tests, we though this to be little artificial as most programmers use JavaScript for web applications.

In this context, we enabled the use of browser debugging tools for stages and the final JavaScript program through a JavaScript wrapper playing the role of the debugger front-end which supports: (i) inspection and visualization of ASTs, with multiple views and code unparsing; and (ii) seamless transition to the next stage with the normal *step* debugging command issued on the last statement of the current stage.

***Contribution*** We present the way a full-scale multi-stage extension of JavaScript is implemented *on-top* of an existing engine, in our case Spider Monkey. No compilation or runtime features are repeated in our implementation, while all staging aspects are *inserted* in a mostly unobtrusive manner. Our application tests are mainly web clients. In this context, we demonstrate generative user-interface implementation through staging, producing comprehensive JavaScript web apps. Finally, we discuss the implementation of the stage-debugging loop in Spider Monkey for debug-mode execution. Also, we explain how we made a JavaScript web client playing the role of the stage-debugging fronted, which together with the browser debugging tools offer the necessary functionality for stage debugging.

## 2. Language Extensions

### 2.1 AST Tags

Such tags are provided to ease AST composition and do not cause staging themselves. Practically, they reify the language parser, the internal AST structures and the AST creation functions.

***Quasi-quotes*** (written **.<***defs***>.**) may be inserted around JavaScript definitions, such as expressions, statements, functions, etc., to convey their AST form and are the easiest way to create ASTs directly from source text. For instance, `.<1+2>.` is the AST for the source text `1+2`. Variables within quasi-quotes are resolved with their names in the context where the respective AST is finally inserted, i.e. are lexically scoped at the insertion point, while name-capture is prevented for identifiers (i.e. hygienic macros) using **$***ident*, making contextually unique identifiers.

***Escape*** (written *.~expr*) is used only within quasi-quotes to prevent converting the source text of *expr* into an AST form by evaluating *expr* normally. The *.~(*expr*)* form, with parenthesis mandated, must be used when escaping a function name. Practically, escape is used on expressions already carrying AST values which need to be combined into an AST constructed via quasi-quotes. For example, assuming *x* already carries the AST value of `.<1>.`, the expression `.<.~x+2>.` evaluates to the AST of `1+2`. The latter also applies in nested quasi-quotes, meaning the expression `.<.<.~x+2>.>.` evaluates to `.<.<1+2>.>.` Additionally, we also support the escaped expression to carry scalar values like number, boolean or string (i.e. *ground* values). In this case, the value is automatically converted to its corresponding AST value as if it were a constant. For instance, if *x* is *1*, then *.~x* within `.<.~x+2>.` will be converted to the AST of value *1*, or `.<1>`, thus `.<.~x+2>.` evaluates to `.<1+2>.`

### 2.2 Staging Tags

Staging tags generally imply preprocessing-time evaluation of associated source code, and are essential in supporting staging. Syntactically, they define the boundaries between stage code fragments and also introduce stage nesting, also known as meta-level shifting.

***Inline*** (written *.!expr*) denotes evaluation of *expr*, whose value must be of AST type, and insertion of its AST value into the enclosing program by replacing itself. The *.!(expr)* form, again with parenthesis required, must be used when in-lining a function name. Inline tags in quasi-quotes are allowed, and, as all other quasi-quoted expressions, are just AST values thus not directly evaluated. This allows expressions carrying an AST with an inline directive to be inlined, meaning inline directives may produce further inline directives, something relating to meta-generators. An example of theoretic value is the self-reproducing staged JavaScript program below, where staging never terminates.

```
function f() { return .< .!f() >.; }
.!f();
```

***Execute*** (written *.&stmt*) defines a stage *stmt* representing any *single statement, local definition or block* in JavaScript. Any definitions introduced are visible only within staged code. Execute tags can also be nested (e.g. *.&.&stmt*), with their nesting depth specifying the stage nesting of the *stmt* that follows. Additionally, execute tags can be quasi-quoted and be converted to AST form, meaning their inlining introduces further staging. A common example used in the literature for staged generators is the power function shown below in our multi-staged JavaScript (exponent is always a natural number), which produces a sequence of multiplications:

```
.& {
   var power = function(base, exp) {
        var res = .< .~base >.;
        for (var i = 0; i < exp; ++i)
            res = .< .~res * .~base >.;
        return res;
   }
}
```

## 3. Staging Implementation

### 3.1 Staging Loop

This loop essentially instantiates the overall staging process. It is invoked exactly after parsing, taking as input the program AST,

```
typedef JSObject* Node;
I:list<Node> (*keeps inline nodes*)

Node GetStage (root: Node) {
  clear I
  Let N: innermost stage nesting
  if N = 0 then (*no more stages*)
     return nil
  Node S = stmtsNode() (*as a JS stmt list*)

  foreach x in root via a DFS traversal
    if x has stage nesting N then {
      append x in S
      if x.type = PNK_METAINLINE then
        append x in I
    }
  return S
}

void StagingLoop (root: Node) {

  Node S = GetStage(root)

  while stage ≠ nil {
    String stageSource = Unparse(S)
    SaveStageSource(stageSource);
    JSEvaluate(stageSource);
    S = GetNextStage(root)
  }
}
```

Figure 3 Outline of the actual staging loop we implemented in Spider Monkey.

which includes the new tags, and results in a modified AST, with no staging-related tags, after staging is evaluated. In the overall translation pipeline, it lies between parsing and code generation, although internally involving code generation and execution rounds of the main JavaScript engine (extracted stages are always pure JavaScript programs). The extracted stage is a standard JavaScript program, which is unparsed and evaluated with a new JavaScript engine instance. The unparsed source is saved to disc for convenience, enabling programmers view the entire stage outside the enclosing program.

Initially, we aimed to adopt the internal AST structures of Spider Monkey (`ParseTreeNode` in C++) and use it throughout the staging process as the input / output data type for ASTs. Then, we observed that using the original AST structures entailed a few issues, including many missing features for tree editing and composition, since `ParseTreeNode` was not designed to be mutable during the translation process. Then, we noticed that Spider Monkey defines the format of JavaScript objects that can be considered as well-formed AST values, offering methods of its *reflection* library to convert between `ParseTreeNode*` and `JSObject`, the latter being Spider Monkey internal representation of JavaScript objects. Thus, we decided to use this facility, and convert the initial AST to `JSObject*`, thus work exclusively on `JSObject*` values as the staging AST representation type.

Based on this, the detailed logic of the staging loop we implemented for Spider Monkey is provided under Figure 3. As shown, when extracting the nodes comprising the current stage, we also keep a list ***I*** of node references, corresponding to the inline directives of the stage with their order in the source. Such references are the actual positions of *inline* tags in the main AST, not the stage one. The list ***I*** helps in evaluating *inline* directives, in particular for replacing their presence in the main AST by their argument, as is discussed later in detail.

### 3.2 Parser Extensions

Not surprisingly, the required extensions are extremely minimal. Firstly, a few new token categories had to be introduced together with very small-scale additions in the scanner. Then, the original AST structures have been extended to accommodate the new required node types. In Spider Monkey, this just meant the insertion of extra values for the `ParseNodeKind` enumerated type (`PNK` also used as a synonym). The original node creation methods were sufficient since all new tags are unary operators with a single AST operand, something also apparent in the following creation expressions (now part of the parser code).

```
new_<UnaryNode>(PNK_METAQUASI,JSOP_NOP,pos,defs)
unaryOpExpr(PNK_METAESC,JSOP_NOP,expr)
new_<UnaryNode>(PNK_METAEXEC,JSOP_NOP,pos,stmt)
unaryOpExpr(PNK_METAINLINE,JSOP_NOP,expr)
```

The only change we had to make in `ParseNode` concerned the addition of an extra union field when using *escaped* or *inlined* function names, as shown below.

```
class ParseNode { …
   union { …
       struct {  …
         ParseNode* escOrInlineFuncName;
       } name;
   } pn_u;
};
```

Finally, a few AST structural assertions had to be loosened by accommodating the presence of staging tags, such as *escapes* and *inlines* as function names and arguments.

### 3.3 AST Conversions

As mentioned earlier, AST tags should be converted to JavaScript objects and vice-versa, adopting in our implementation the Spider Monkey AST objet format of its *reflection* library. In the example below we show code composition at the level of ASTs in staged code (*execute* tag), using quasi-quotes and *escapes*, without yet causing any code generation (since no *inline* is defined).

```
.& {
 function Ctor (name, args, stmts) {
    return .<
         function .~(name)(.~args) {
               .~stmts;
          }
    >.;
 }
}

a = .< x, y >.;
s = .< this.x = x; this.y = y; >.;

point2dCtor = Ctor (.< Point2d >., a, s);
point3dCtor = Ctor (
  .< Point3d >.,
  .< .~a, z >.,
  .< .~s; this.z = z; >.
);
}
```

The previous stage is translated to a standard JavaScript program, with the `Ctor` function provided in Figure 4. As shown, quasi-quotes are translated to nested object construction expressions. The details of the Spider Monkey AST format can be observed in the nesting structure, element arrays and actual field keys. The

```
function Ctor (name, args, stmts) {
     return {
         type: "Program",
         body: [{
            type: "FunctionDeclaration",
            id: meta_escape(
                  "SINGLE_ELEM",
                  name,
                  "IS_EXPR"
                ),
            params: meta_escape(
                      "LIST_ELEM",
                      [],
                      [{index: 0, expr: args}],
                      "IS_EXPR"
                    ),
            body : [{
                type: "BlockStatement",
                body: meta_escape(
                      "LIST_ELEM",
                      [],
                      [{index:0, expr: stmts}],
                      "IS_STMT"
                    )
            }]
        }]
     };
 }
```

Figure 4 Translation in JavaScript of the quasi-quotes and escapes of the `Ctor` function of Section 3.3.

```
void MetaInline (Node ast) {
  Node inline = I.front()
  ReplaceInParent(inline, ast)
  I.pop_front()
}

Node MetaEscape (
   Node     ast,      (*the escaped tree*)
   Node     parent,   (*the enclosing context*)
   Position pos       (*placement information*)
) {
  if pos.isListElem = true then
    if pos.isStmt = true then
      InsertInStmtList(parent, ast, pos.index)
    else
      InsertInExprList(parent, ast, pos.index)
  else
  if pos.isStmt = true then
    InsertSingleStmt(parent, ast)
  else
    InsertSingleExpr(parent, ast)
  return parent;
}
```

Figure 5 Outline of the implementation of *inline* and *escape* library functions in Spider Monkey.

three cases of *escape* use in quasi-quote, regarding the function name, arguments and statements, result in respective invocations of the `meta_escape` library function performing the required concatenations when the stage is evaluated.

As said earlier, although the `Ctor` is defined in a stage, its implementation involves no staged elements as such, since the special AST tags for AST manipulation have non-staged semantics. In this sense, they may well be provided by a non-staged extension of standard JavaScript offering quasi-quotes, escaping and AST composition for improved reflection. We discuss in the next section the details of their implementation.

### 3.4 Code Insertion ( *inline* and *escape* Tags)

A single stage program may encompass multiple *inline* tags, placed as needed at various locations of the enclosing program, with their role to compose and insert source code fragments. The particular way such tags are arranged in the source code is actually a matter of the generative metaprogramming approach that the client programmer aims to implement.

The handling of *inline* tags is straightforward. As mentioned earlier, stages become pure JavaScript programs where the *inline* code-insertion tag is converted to the invocation of a respective library function, installed upon start-up on the extended Spider Monkey engine. In its implementation is `MetaInline`, at top of Figure 5, it accesses and modifies the list of inline node references ***I*** of Figure 3. The sequence of inline nods in ***I*** corresponds to the exact sequence of *inline* calls as met in the stage program. The argument to `MetaInline` is a `JSObject*` (`Node` as synonym) carrying the AST of the code fragment to be inserted. Extending the previous stage example, where we composed two constructor ASTs carried in the `point2dCtor` and `point3dCtor` variables, we add the following extra four lines:

```
.!point2dCtor; (*staged code, 1st inline in I*)
var pt2d = new Point2d(10,20);
```

```
.!point3dCtor; (*staged code, 2nd inline in I*)
var pt3d = new Point3d(10, 20, 30);
```

In this case, the two *inline* tags are concatenated exactly after the code block of the *execute* tag (.&) of our earlier example, making the first stage to evaluate, being also the only stage in this case. The stage assembly logic has been outlined earlier, in the staging loop under Figure 3. As discussed, it collects all statements at the same stage nesting, in the order they are met, and, inserts inline tags belonging to the stage in the special ***I*** list. Thus, for the example, ***I*** contains the node references for the expressions `.!point2dCtor` and `.!point3dCtor` from the AST of the main program.

The evaluation of the two *inline* tags causes a rewriting of the main AST by replacing the *inline* directives with the content of *point2dCtor* and *point3dCtor*. It results in the following final code of the main program, after staging is performed.

```
function Point2d(x,y) {
    this.x = x;
    this.y = y;
}

var pt2d = new Point2d(10, 20);

function Point3d(x,y,z) {
    this.x = x;
    this.y = y;
    this.z = z;
}

var pt3d = new Point3d(10, 20, 30);
```

The implementation of the *escape* tag, `MetaEscape` function at bottom of Figure 5, requires the parent node and palcement information, within the surrounding quasi-quoted AST. The `parent` node is either a single *expr* / *stmt*, thus `ast` argument is hooked as a single child, or a list of *expr* / *stmt* nodes (internally an array), with the `ast` inserted exactly at `pos.index` location.

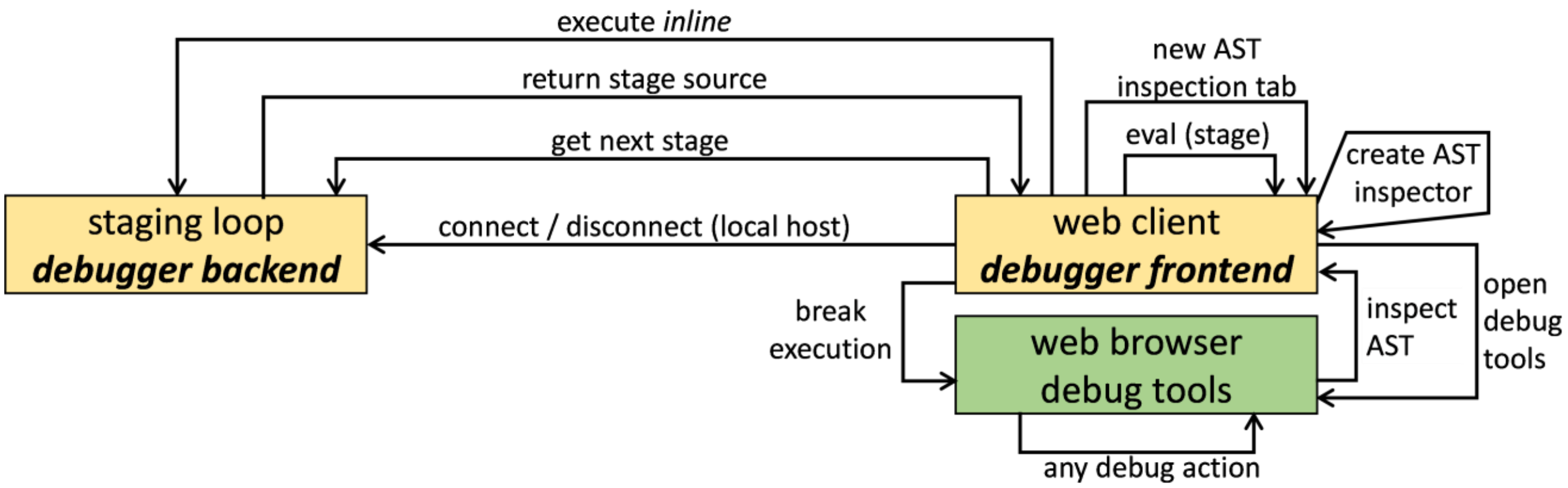

Figure 7 Stage debugging architecture by splitting responsibilities between staging loop for stage extraction, the web browser tools for typical debugging activities, and a custom web-client for stage evaluation and AST inspection.

It should be noted that `parent` and `pos` arguments can be dropped by an alternative implementation checking the parent node of `ast` inside `MetaEscape` to identify the insertion method. We decided to avoid this processing since context information is already available during the syntactic processing of *escapes*.

## 4. Debugging Support

To support debugging of stages we have implemented a back-end service loop as part of the staging loop, which basically responds in two requests (besides the apparent connect / disconnect ones): (i) extract and return the next stage; and (ii) apply an inline directive. The escape directive affects locally the ASTs composed in a stage and was implemented as a JavaScript function merged with the produced stage source. With this we avoided an extra message, something though not possible with *inlines* that affect the main AST that is kept in the running Spider Monkey instance.

For the debugger front-end we made a web client that communicates with the back end, with the overall architecture, messages and actions outlined under Figure 7. Initially, the client opens the debugging tools and creates a window where visual AST inspection is supported with multiple tabs. The primary client-debugging loop is provided below.

```
function startDebugging() {

  var onNextStageStart = function(msg) {
    if (msg.stage == 0) // was the last stage
       openPage('Finished stage debugging');
   else {
     var src = unescape(msg.src);
     openPage('Debugging stage', msg.stage);
     setOnLoadedPage(
        function() { // client debug loop
          eval(src); // execute current stage
          startDebugging(); // debug next one
        }
     );
    }
  }

  StageDebugSend({ // backend-frontend comm
    header:    'Next',
    async:     true,
    success:   onNextStageStart, // on receive
    fail:      onNextStageError
  });
}

openPage(
  'Starting stage debugging',
   startDebugging
);
```

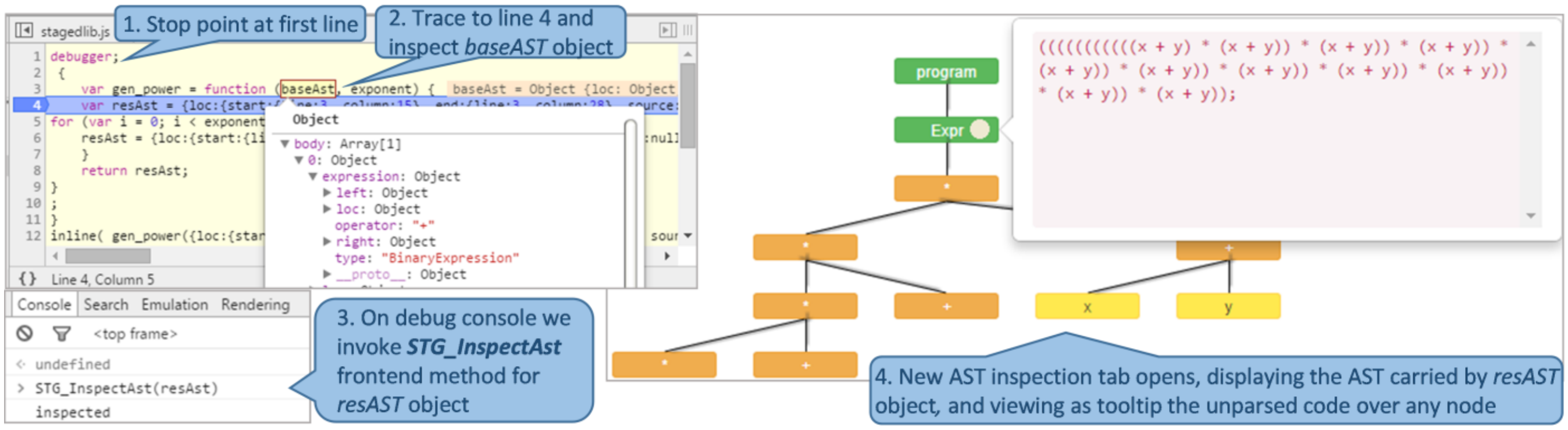

Figure 6 Stage debugging session with the initial stop point, custom tracing and inspection with debug tools, and activation of AST visual inspection offered by the frontend using the debug console.

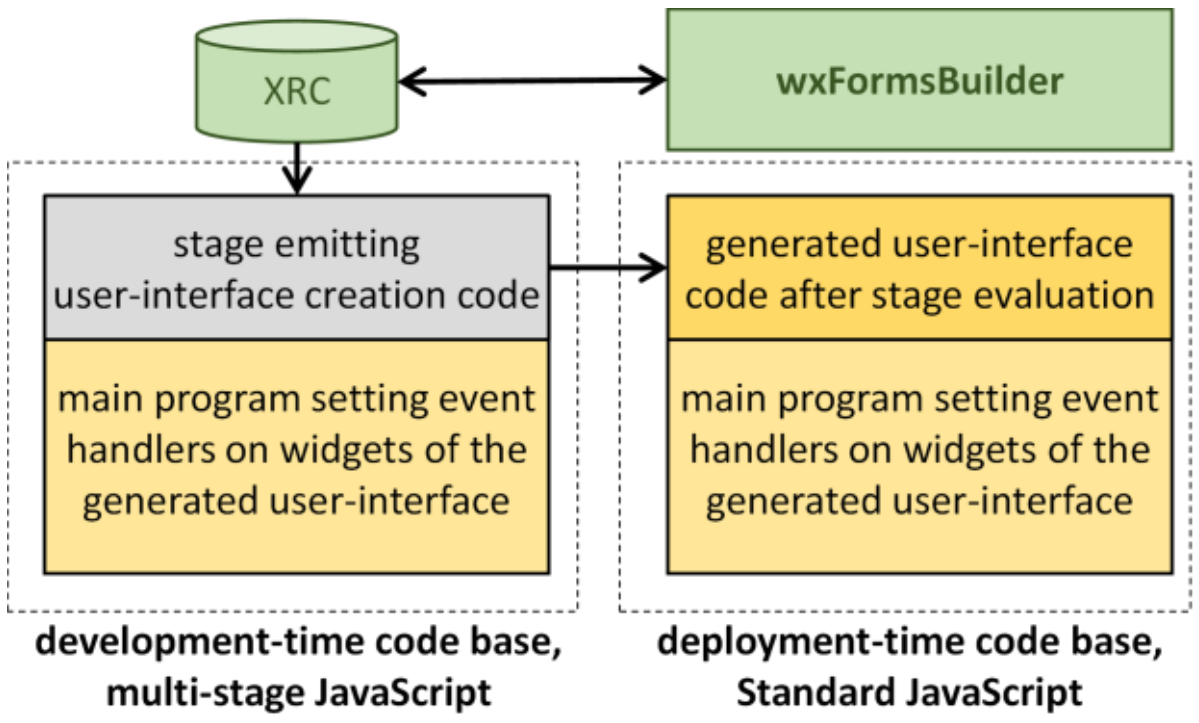

Figure 8 Toolchain for staged code generation of JavaScript web user-interfaces from the output of an interface builder.

As shown, the client loop it is not actually implemented as an explicit loop, but repeats stage debug sessions as follows: (i) a message is sent to the staging loop backend requesting to extract the next stage, while also setting a local response handler function `onNextStageStart`; (ii) on receipt of the response, the handler function checks if there are no more stages (`msg.stage==0`), else it removes character escape sequences from the stage source and directly evaluates it by restarting a new debug round. In order to support stage tracing with the browser debugging tools, the staging loop inserts a `debugger;` statement directly at the first line of *every* stage source posted to the frontend client.

The latter causes an instant break point when `eval(src)` is invoked, stopping execution with the debugger tools at the first line, before the stage is essentially executed. At this point, additional breakpoints may be interactively inserted by the programmer as desired, using the debug tools. An example debug session for the *power* staged function (mentioned in Section 2.1) is shown under Figure 6.

## 5. Application Example

There are many examples regarding the use of generative macros in the literature that we have also used in our previous work on multi-stage languages [5]. But since JavaScript is mostly deployed as a web programming language we decided to practice generative stage-based programming directly on this domain. In this context, it easy is to observe that code generation is widely applied in so called dynamic web applications, where HTML and JavaScript

```
.& {
   var XRC_PATH = "<UI builder output path>";
   var xrcElems =  LoadSpecs(XRC_PATH);

   // generators make ASTs for: creating
   // widgets and setting their attributes
   // form the data carried in xrcElem
   function GenerateButton (xrcElem)
         {…}
   function GenerateStaticText (xrcElem)
         {…}
   function GenerateChoice (xrcElem)
         {…}
   // rest of generator functions here

   var dispatcher = {
        'Button'         : GenerateButton,
        'StaticText'     : GenerateStaticText,
        'Choice'         : GenerateChoice,
        // rest of generators installed here
   };

   function GenerateUI() {
     var ast = nil;
     for (var e in xrcElems) {
       var curr = dispatcher(e.elemClass)(e);
       ast = .< .~ast; .~curr; >.; // compose
     }
     return ast; // AST of UI creation code
   }
}

.!GenerateUI();
```

Figure 9 Outline of the stage-based web-UI generator from XRC specifications.

code results from server-side processing in a component commonly referred as the *Presentation Tier* in *n-tier* architectures. Currently, the trend of Rich Internet Applications (RIAs) emphasizes comprehensive user-interfaces for web-applications directly in JavaScript, by minimizing inherent page changes when refreshing content, and suggesting the local handling of such required updates at the client side, thus minimizing overall page reloading.

Based on these remarks, we started thinking to generate JavaScript user-interface source code using exclusively stages, considering this to be a fresh but also a quite demanding application scenario for a multi-stage JavaScript. In particular, we implemented the following application scenario.

We used a public graphical interface builder (i.e., a rapid prototyping tool), named *wxFormsBuilder*, which supports the visual design of user-interfaces, and outputs an XML-based file in a format named XRC. The latter describes the widget hierarchy, placement and layout constraints, and visual attributes. Then, we realised the staging-driven toolchain of Figure 8, to craft the user-interface of a lightweight e-mail client, as shown under Figure 10. It should be noted that the staging process is meant only for development-time, since it relies on no deployment-time parameters neither it interoperates with the browser host environment (i.e., DOM). The latter puts the staged code as the development-time code-base, with the finally produced code used only for deployment, not for editing or evolution, since it is overwritten after every recompilation.

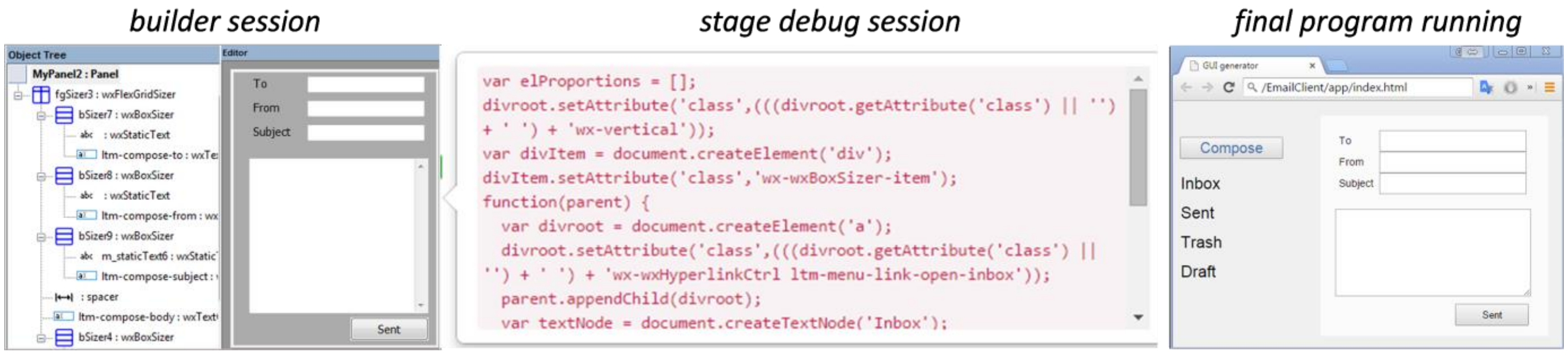

Figure 10 Part of an actual development cycle, in an example mail client, involving a builder session, a stage debug session and a test session with the final program running in the browser.

The source code for realizing this generative user-interface implementation approach is provided under Figure 9. It is basically split in three parts: (i) loading of specifications into an array of objects carrying element creation information; (ii) a set of generator functions composing ASTs for the respective widget creation code, including statements which apply the visual attributes carried in the element argument; and (iii) a loop iterating on all elements and invoking generators while compositing the final AST that is eventually *inlined* to flush the respective code inside the program. The implementation size of the staged generator is about 350 lines, and remains constant in relationship to the target user-interface made with the builder and the final XRC specifications. However, the generated user-interface in JavaScript code for our e-mail client is around 1000 lines, something demonstrating the significant potential of staged generators.

Finally, it should be mentioned that we realized how crucial stage debugging is, and invested in supporting it, only during the development of the e-mail client, since it resulted in a stage quite bigger than then tens of lines of code of our previous examples (really, nobody needs a stage debugger to make a correct staged *power* function).

## 6. Related Work

There has been a considerable amount of work in the literature regarding multi-stage languages and macro systems, with a few also focusing on JavaScript. As mentioned earlier, we do not argue a contribution in the context of multi-stage language elements or metaprogramming models in general, or for JavaScript in particular. Our approach was an arguably successful trial for an extension-based implementation style of staging on top of an existing JavaScript interpreter, by fully reusing all its features. For example, the Spider Monkey JIT features work just fine when stages are evaluated. In this context, we review previous work to show the types of multi-stage implementations available and compare with respect to our engineering style.

In [2], a Scheme-like macro system for JavaScript is discussed, emphasizing a hygienic approach for macros particularly suited to JavaScript. In this implementation macros are not JavaScript code, but a custom language, with a new processing pipeline on its own. While macro hygienic is not a focus of our work, we should mention that we addressed it with selective automatic renaming, as in our work in [5], by tracing conflicting identifiers in the surrounding lexical scope *using* solely the AST, involving no lexical analysis issues.

Template Haskell [9] is a two-stage language that provides metaprogramming facilities through quasi-quotes and splicing (same as *inline* tags). It reuses most aspects of the normal language without however providing debugging support for stages.

*Converge* [12][13] is a dynamic class-based language that allows CTMP in the spirit of Template Haskell. In principle, its language staging layer fully reuses the normal language features, compiler and execution system, building essentially on top of it, although there is no distinction between the two layers since the language was designed from the beginning to be staged.

Metalua [4] is a compile-time metaprogramming extension of the highly popular untyped Lua language, and supports stages with the concept of separated meta-levels, allowing shifting between them using special syntax. Although the original language constructs are fully available in stages, its implementation features a separate scanner and parser, and a custom reimplementation of its virtual machine for stage evaluation.

Groovy [10] supports compile-time transformational metaprograms through AST transformations written in Groovy itself. Thus, it reuses the language compiler and runtime system, while it is possible to debug local transformations directly from the IDE. Compared to previous languages, its programming model is far from typical macro systems, and is single staged since there is no notion of meta-transformations (i.e., transformations generating transformations). However, as with Converge, it is implemented in a modular manner on top of the normal language layer.

The two major Lisp dialects, Common Lisp [7] and Scheme [3][3], support metaprogramming through their powerful macro systems. In Common Lisp, programs can manipulate source code as a data structure, while Scheme macros are transformation procedures accompanied by a simple pattern matching sub-language. At the implementation level both languages have a single interpreter for macros and the rest of their features. Now, judging from the previous mentioned languages, it seems that such sharing and reuse is a common pattern when a language is designed to originally support macros, compared to adding staging as an extension feature far later.

## 7. Summary and Conclusions

We have implemented[1] a multi-stage version of JavaScript on top of the Sider Monkey engine by fully reusing the original parser and runtime, introducing only extra functionality required to support staging. Our overall conclusion is that building multi-stage extensions on top of original implementations can be relatively straightforward and modular, with no wheels reinvented. In the staging extensions we also included the *execute* tag [5], making more convenient the composition of comprehensive stage programs. We also focused on larger-scale application scenarios, being more close to the common deployment of the JavaScript language, compared to smaller-scale macro-related examples appearing in the literature. In this context, we applied generative meta-programs to emit the source code of web clients entirely in JavaScript. Such application tests, due to their relative complexity, required stage debugging and visual AST inspection with code unparsing. This led us to support stage debugging by fully reusing browser debug tools, and adding only the AST-specific facilities (also no wheels reinvented).

Overall, we managed to realise a full-scale multi-stage extension, yet with minimal intervention and full reuse of the underlying JavaScript engine. We also managed to preserve the same principle for the debugging support as well. In the meantime, we explored more demanding scenarios to assess multi-stage features, by practicing generative application-level development.

We believe that with this work we also demonstrate the development cost for generative metaprogramming features in JavaScript to be low and manageable.

[1] https://github.com/apostolidhs/MetaMonkey